\documentclass[aps,prl,twocolumn,letter,longbibliography,superscriptaddress]{revtex4-1}
\usepackage[english]{babel}

\usepackage{graphicx}
\DeclareGraphicsExtensions{.pdf}

\usepackage{float}
\usepackage{amsmath}
\usepackage{color}
\usepackage{ulem}

\newcommand{\be}{\begin{equation}}
\newcommand{\ee}{\end{equation}}

\definecolor{ceccogreen}{rgb}{0.1, 0.8, 0.1}

\begin{document}
\title{Marginal Stability Enables Memory Training in Jammed Solids}
\author{Francesco Arceri}
\affiliation{Department of Physics, University of Oregon, Eugene, Oregon 97403, USA}
\author{Eric I. Corwin}
\affiliation{Department of Physics, University of Oregon, Eugene, Oregon 97403, USA}
\author{Varda F. Hagh}
\affiliation{Department of Physics, University of Oregon, Eugene, Oregon 97403, USA}
\affiliation{James Franck Institute, University of Chicago, Chicago, IL 60637, USA}
\date{\today}

\begin{abstract}
Memory encoding by cyclic shear is a reliable process to store information in jammed solids, yet its underlying mechanism and its connection to the amorphous structure are not fully understood. When a jammed sphere packing is repeatedly sheared with cycles of the same strain amplitude, it optimizes its mechanical response to the cyclic driving and stores a memory of it. We study memory by cyclic shear training as a function of the underlying stability of the amorphous structure in marginally stable and highly stable packings, the latter produced by minimizing the potential energy using both positional and radial degrees of freedom. We find that jammed solids need to be marginally stable in order to store a memory by cyclic shear. In particular, highly stable packings store memories only after overcoming brittle yielding and the cyclic shear training takes place in the shear band, a region which we show to be marginally stable.
\end{abstract}

\maketitle

\textit{Introduction} -- 
When subject to a repeated driving, amorphous solids are able to adapt their spatial structure to the external deformation~\cite{corte_random_2008}. By doing so, they store a memory of the periodic driving as a structural information which can be later extracted~\cite{fiocco_encoding_2014, keim_memory_2019}. A widely used protocol for encoding a memory in jammed solids is cyclic shear training~\cite{corte_random_2008, fiocco_oscillatory_2013}: the system is repeatedly sheared with cycles of strain amplitude $\gamma_{train}$, until it reaches a periodic orbit, i.e. a sequence of rearrangements that the system undergoes every time the same cyclic perturbation is applied. Cyclic shear training finds an explanation in the complex energy landscape of amorphous solids~\cite{charbonneau_fractal_2014-4, jin_exploring_2017} where each rearrangement corresponds to a transition between two energy minima. As the training goes on, the system finds the most energetically favorable path between minima optimizing the mechanical response to the cyclic deformation~\cite{fiocco_oscillatory_2013, szulc_forced_2020}. While previous studies have shown that cyclic shear brings the system to a lower energy minimum~\cite{adhikari_memory_2018}, recent advances in producing extremely annealed glassy configurations in thermal~\cite{ninarello_models_2017} and athermal simulations~\cite{kapteijns_fast_2019} have led to the conclusion that the rheology of amorphous solids is ruled by the preparation protocol~\cite{rainone_following_2016-1, jin_stability-reversibility_2018,  yeh_glass_2020}. In particular, cyclic shear is only efficient in lowering the energy of marginally stable glasses~\cite{adhikari_memory_2018, yeh_glass_2020}, i.e. configurations that become unstable under very small perturbations~\cite{biroli_breakdown_2016-1, charbonneau_glass_2017-3}. By contrast, cyclic shear fails to further anneal highly stable glassy configurations~\cite{yeh_glass_2020}. Here, we explore the connection between memory training by cyclic shear and mechanical stability in jammed solids and show that memory training is only possible when the system, or a portion of it, is marginally stable.

We produce highly stable packings of jammed soft spheres via a recently developed algorithm based on the simultaneous minimization of positional and radial degrees of freedom~\cite{hagh_transient_2021}, while a conventional FIRE minimization is used to produce marginally stable packings~\cite{durian_foam_1995-1, ohern_jamming_2003-3}. While marginally stable packings show ductile behavior upon increasing the applied shear strain~\cite{maloney_amorphous_2006-2}, highly stable packings are brittle and yield by forming a shear band~\cite{maloney_universal_2004, rainone_following_2016-1, ozawa_random_2018, kapteijns_fast_2019}. Subject to cyclic shear training, marginally stable packings store memories down to low strain amplitudes and show a uniform participation to the training. By contrast, highly stable packings can only store memories past the yielding strain and only the particles in the shear band actively participate to the training. Here we show that the shear band is a marginally stable region of the system and its size controls the memory training.

\begin{figure} [b]
\includegraphics[width=0.9\columnwidth]{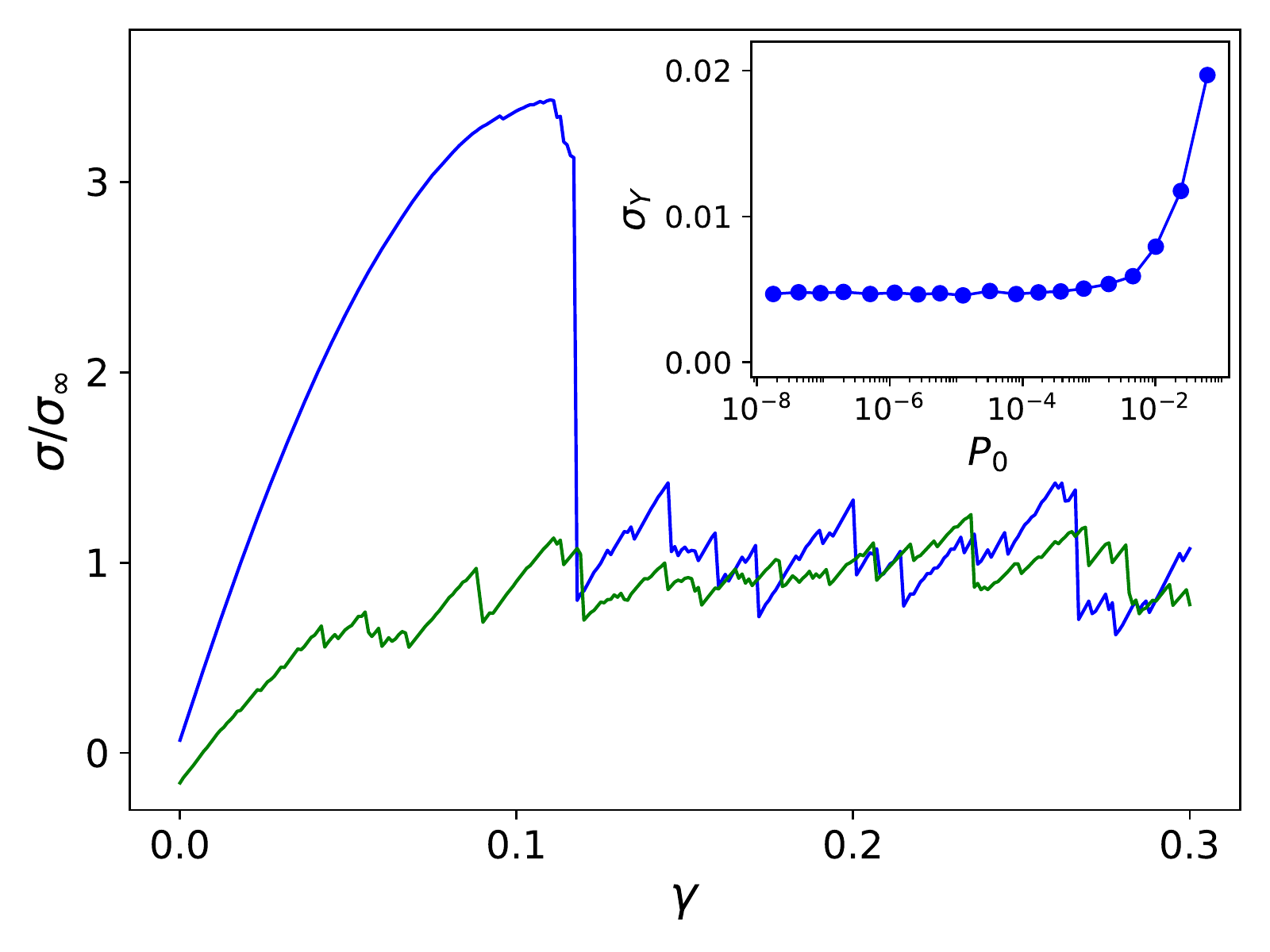}
\caption{Stress vs. strain curves for highly stable (blue) and marginally stable (green) packings produced at pressure $P_0 \simeq 0.08$ and composed of $N=4096$ particles. The stress is scaled by its typical value $\sigma_\infty$ in the plastic regime after yielding. Inset: yielding stress, $\sigma_Y$, as a function of the initial pressure, $P_0$, at which brittle packings are produced.}
\label{fig:linear-shear} 
\end{figure}

\textit{Numerical methods} -- 
We produce samples of athermal soft sphere packings using the \texttt{pyCudaPacking} package developed by Corwin \textit{et al.}~\cite{morse_geometric_2014-5, charbonneau_jamming_2015-6}. Each packing is composed of $N$ particles contained in a three dimensional simulation box of unitary volume with periodic boundary conditions. Particles interact via the soft sphere harmonic contact potential 
\begin{equation}
U_{ij} = q_{ij}^2\Theta(q_{ij}) \; \mbox{,} \;\;\;\;\; q_{ij} = 1 - \frac{|\vec{r}_{ij}|}{\sigma_{ij}}
\end{equation}
where $\vec{r}_{ij}$ is the distance between particles $i$ and $j$, $\sigma_{ij}$ is the sum of their radii, and $\Theta$ is the Heaviside step function. We use a log-normal distribution of particle sizes with $20\%$  polydispersity to avoid nucleation of crystalline structures. This model undergoes the jamming transition at zero pressure where particles share just enough contacts to enforce global rigidity~\cite{ohern_jamming_2003-3}. We produce marginally stable packings by minimizing the energy with respect to only positional degrees of freedom via the FIRE algorithm~\cite{bitzek_structural_2006-2}. To produce highly stable packings, we add particle radii as constrained variables to the minimization. In particular, we start from a configuration with random positions and polydisperse size distribution, and allow both particle positions and radii to relax in order to minimize the energy. To keep the initial size distribution fixed, we constrain the radial components of the particle forces by fixing a set of moments of the distribution, namely $\{-6, -3, 3\}$~\cite{hagh_transient_2021}. Once the energy is minimized, we fix the radii and perform the shear training.

\begin{figure}
\includegraphics[width=\columnwidth]{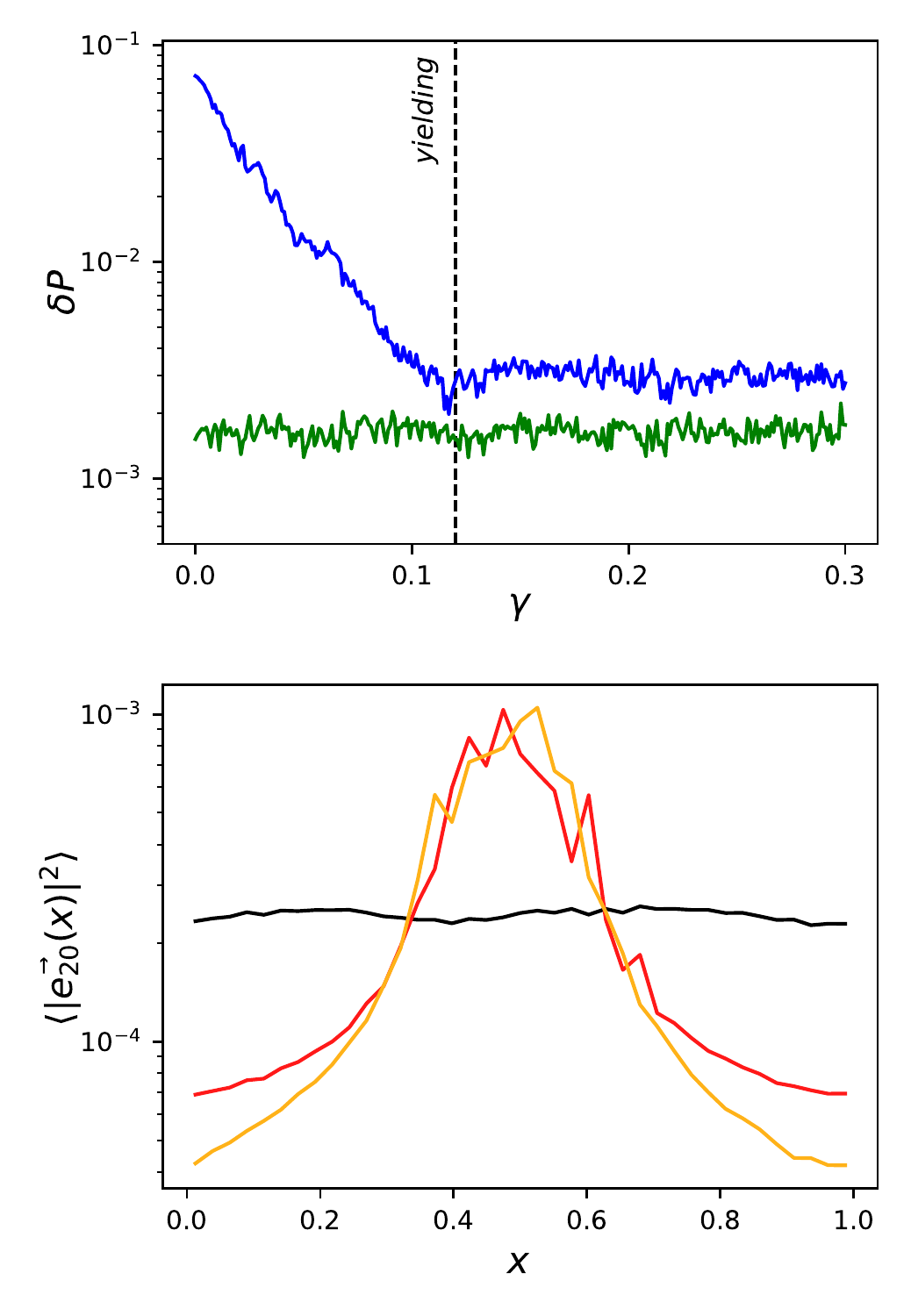}
\caption{Top: pressure change, $\delta P$, required to push a packing to a nearby instability as a function of the applied strain $\gamma$ averaged over $20$ samples for both highly stable (blue) and marginally stable (green) packings of $N=4096$ particles. The dotted line indicates the average yielding strain of highly stable packings. Bottom: magnitudes of the first $20$ low-frequency eigenvectors, averaged over slices of the system along the $x$ axis at zero strain (black), right after yielding at $\gamma=0.122$ (red), and in the plastic regime at $\gamma=0.3$ (yellow) for a highly stable packing. Data are shifted to center the shear band.}
\label{fig:pressure} 
\end{figure}

We simulate athermal quasistatic shear (AQS) along the $yx$ direction by applying steps of $\Delta\gamma = 10^{-3}$ strain with Lees-Edwards boundary conditions. A single strain step consists of an affine displacement of each particle $(x_i, y_i, z_i) \rightarrow (x_i, y_i + \Delta \gamma x_i, z_i)$, followed by a minimization of the potential energy with respect to the positional degrees of freedom only. We choose to study configurations produced at pressure $P_0 \simeq 0.08$ to optimize the computational cost of our simulations which slows down as the jamming transition is approached.

The rheology of marginally stable and highly stable packings is depicted in Fig.~\ref{fig:linear-shear}. Marginally stable packings show ductile behavior as they encounter the first instability at very small strain and yield through a series of plastic rearrangements~\cite{maloney_amorphous_2006-2}. On the other hand, highly stable packings are brittle: they show an elastic response up to a large yielding strain, $\gamma_Y$. After yielding, a sharp stress drop signals the failure under the external load and the system breaks along a shear band~\cite{maloney_universal_2004, kapteijns_fast_2019}. In the inset of Fig.~\ref{fig:linear-shear}, we plot the yielding stress, $\sigma_Y$, of highly stable packings as a function of the pressure at which they are produced, $P_0$. The yielding stress plateaus to a finite value as the jamming point is approached in the limit $P_0 \rightarrow 0$ showing that highly stable packings are brittle down to extremely low pressures.

\textit{Evolution of stability under shear} -- 
To understand the relation between the mechanical stability of a packing and its ability to store memories of shear amplitudes, we first study the evolution of mechanical stability upon increasing the applied shear strain. Before reaching the yielding transition, highly stable packings are characterized by a smooth rise of both pressure and energy in the elastic regime. At the same time, the low-frequency vibrational density of states, which rules the linear response of the system~\cite{manning_vibrational_2011-2}, is progressively shifted towards lower frequencies. These properties suggest that highly stable packings would become unstable under an increasingly smaller perturbation as they approach the yielding point. We investigate how the stability of a packing evolves during AQS by computing the change in pressure, $\delta P$, required to push the system to an instability without changing the contact network~\cite{hagh_transient_2021}, as reported in the top panel of Fig.~\ref{fig:pressure}. In marginally stable packings, the distance to a nearby instability fluctuates around a typical value across all the explored range of strain. Highly stable packings present a very different behavior. At zero strain, they require a large change in pressure to find a nearby instability. As the system is progressively sheared, $\delta P$ decreases following an exponential decay which ends at the yielding point. After yielding, $\delta P$ follows a similar behavior as for marginally stable packings. The behavior of $\delta P$ implies that highly stable packings lose stability and become marginally stable after yielding. 

We then explore how the progressive loss of stability in highly stable packings influences the spatial structure of the system by computing the first $20$ low-frequency eigenvectors of the Hessian, i.e. the vibrational modes which control the particle motion under small perturbations~\cite{manning_vibrational_2011-2}. In the bottom panel of Fig.~\ref{fig:pressure}, we report the averaged magnitude of the low-frequency eigenvectors as a function of the applied strain. At zero strain, the motion due to small perturbations spans the entire system uniformly, a typical behavior for highly stable jammed solids~\cite{mizuno_continuum_2017}. After yielding, the motion of the low-frequency eigenvectors stays confined in the shear band while the rest of the system is less susceptible to external perturbations. The shear band is then a marginally stable region of the system where particles are more likely to rearrange under quasistatic deformations. We can now show that the existence of a shear band in highly stable packings past yielding is necessary for training a memory by cyclic shear.

\begin{figure}
\includegraphics[width=0.95\columnwidth]{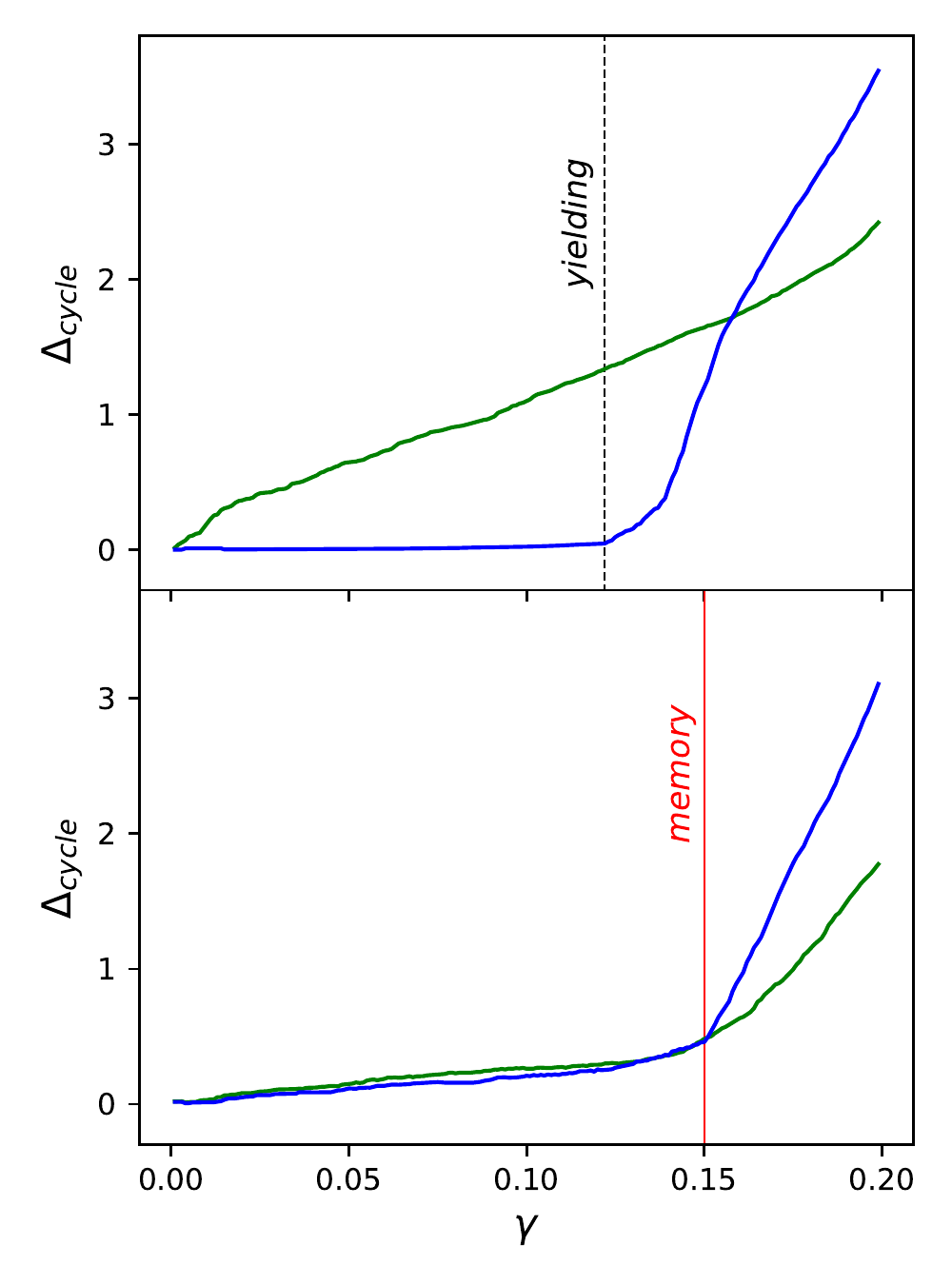}
\caption{Readout shear: $\Delta_{cycle}$ as a function of the strain amplitude $\gamma$ for untrained (top) and trained (bottom) configurations of highly stable (blue) and marginally stable (green) packings. The solid red line indicates the encoded strain amplitude, $\gamma_{train}=0.15$, and the dashed black line shows the average yielding strain for highly stable packings.}
\label{fig:readout} 
\end{figure}

\textit{Memory training} -- 
We use AQS to encode a memory of a strain amplitude, $\gamma_{train}$, by cyclic shear in both marginally and highly stable packings. The following results represent averages over $35$ samples of $N=1024$ particles for both cases. We train a packing by repeating shear cycles until the system reaches a periodic orbit which we identify when the energy at the end of a cycle does not change after one or more consecutive cycles. The encoded memory can then be extracted using a readout~\cite{keim_generic_2011, fiocco_encoding_2014}: starting from a configuration at zero strain, we perform a cycle of strain amplitude $\gamma$ and measure the distance between the initial and final configurations as
\begin{equation}
\Delta_{cycle} = \sqrt{\sum_i |\vec{r_i}^{final} - \vec{r_i}^{initial}|^2} \;\;\; ,
\end{equation}
where the sum runs over the stable particles, i.e. those with at least $d+1$ force bearing contacts~\cite{goodrich_finite-size_2012-5}. 
The readout is performed for a range of strain amplitudes $\gamma \in [0,0.2]$, separated by an increment of $\Delta\gamma=10^{-3}$.

Before training a memory, the readout plots for marginally and highly stable packings show two very different behaviors, as can be seen from the top panel of Fig.~\ref{fig:readout}. The readout for highly stable packings (blue) shows that these are reversible for any cycles with $\gamma < \gamma_Y$ as $\Delta_{cycle}$ stays equal to zero up to the yielding transition. After the brittle failure, $\Delta_{cycle}$ shows an upturn and the system becomes irreversible. For marginally stable packings (green), by contrast, $\Delta_{cycle}$ monotonically increases from zero starting at the beginning of the readout. This indicates that a marginally stable packing undergoes irreversible rearrangements for all the explored strain amplitudes. The readout plots for trained packings is shown in the bottom panel of Fig.~\ref{fig:readout}. Here, $\Delta_{cycle}$ stays close to zero for cycles of strain amplitudes smaller than the training strain, $\gamma_{train}=0.15$. Note that $\gamma_{train}$ is larger than the average yielding strain of highly stable packings. For $\gamma > \gamma_{train}$, both plots show a quick upturn, which is a signature of the memory encoded by cyclic shear training.

\begin{figure}
\includegraphics[width=0.98\columnwidth]{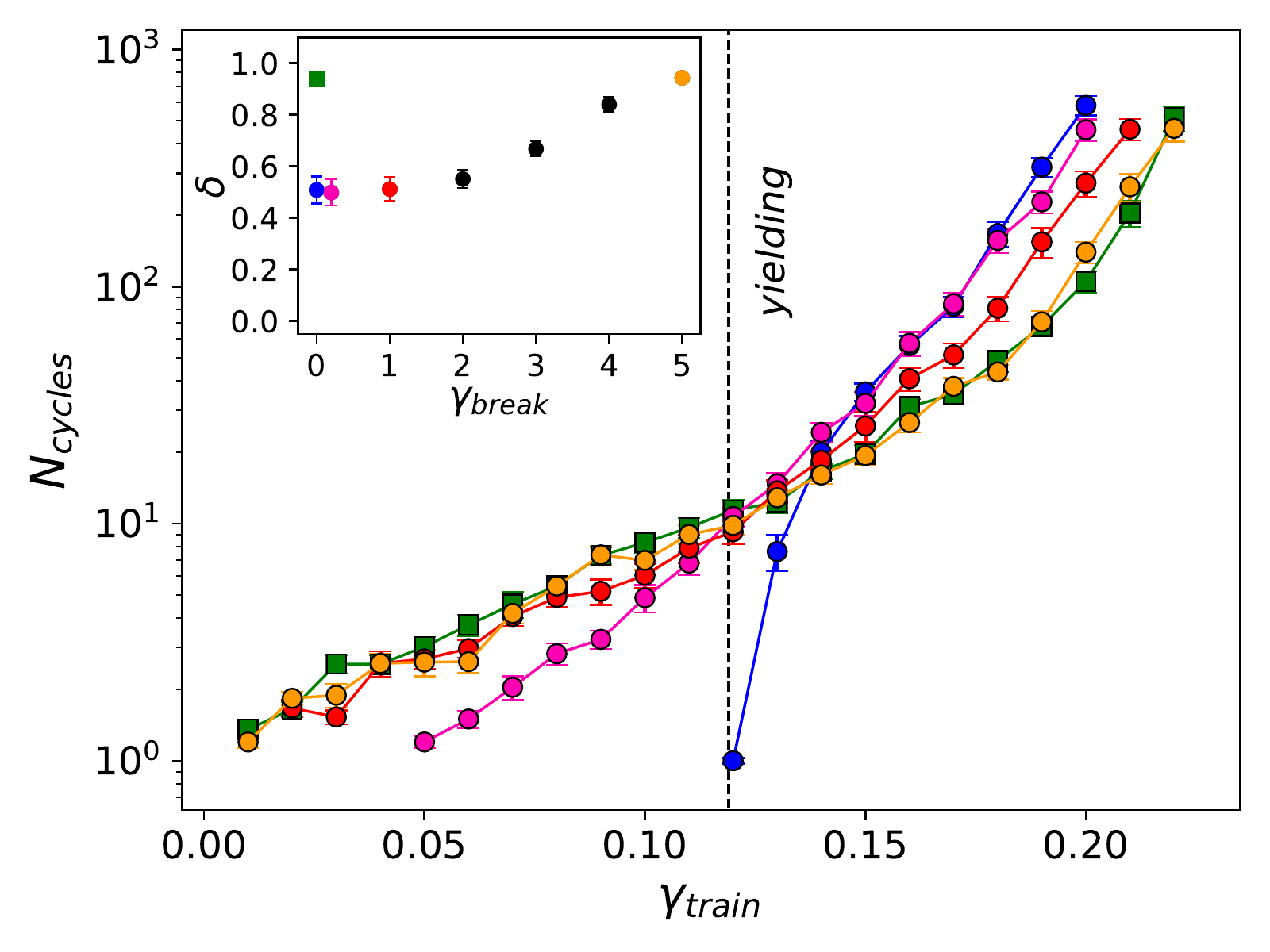}
\caption{Number of training cycles required to encode a memory, $N_{cycles}$, as a function of the encoded strain amplitude $\gamma_{train}$ for brittle (blue) and ductile (green) packings as well as brittle packings that are broken with a single cycle of strain amplitude of $\gamma_{break}=0.2$ (pink), $1$ (red) and $5$ (yellow) before the training. Inset: shear band size, $\delta$, as a function of $\gamma_{break}$ after training a memory of $\gamma_{train}=0.15$ with the same color code as in the main plot. The error bars represent the standard error on the mean.}
\label{fig:limit-cycles} 
\end{figure}

We study the trainability of our packings by plotting the number of training cycles, $N_{cycles}$, needed to encode a memory as a function of the training strain amplitude, $\gamma_{train}$, see Fig.~\ref{fig:limit-cycles}. While marginally stable packings store memories for all the explored ranges of $\gamma_{train}$, highly stable packings are able to store memories only for strain amplitudes larger than the yielding strain. Moreover, at a fixed strain amplitude, highly stable packings need a larger $N_{cycles}$ to reach a periodic orbit compared to marginally stable packings. This is due to the difference in the fraction of particles which are actively participating to the training: while in marginally stable packings all the particles are uniformly displaced by the shear cycles, in highly stable packings the particles within the shear band rearrange much more than others.

To support this claim, we study the relation between marginal stability and the number of training cycles by tuning the width of the shear band. This is accomplished by shearing brittle packings with an initial cycle of large strain amplitude, $\gamma_{break}$, before performing cyclic shear training at a given $\gamma_{train}$. During the breaking cycle, particles adjacent to the shear band relax and lose their initial stability. We estimate the size of the shear band, $\delta$, by computing the distribution of $\Delta_{cycle}$ along one of the transverse directions to shear and extracting the width of the distribution peak. We measure $\delta$ both after the initial breaking cycle and cyclic shear training and find it to be the same within error. As shown in the inset of Fig.~\ref{fig:limit-cycles}, the size of the shear band computed after training a memory of $\gamma_{train}=0.15$ is proportional to $\gamma_{break}$. Fig.~\ref{fig:limit-cycles} shows that for any $\gamma_{break}$, broken brittle packings are able to store memories of strain amplitudes below the yielding strain. As $\gamma_{break}$ increases, the trainability curve gets closer to the one for ductile packings (green). For $\gamma_{break}=5$ (yellow), the shear band is spread out to the entire system and the number of training cycles for strain amplitudes above the yielding strain are similar to those reported for ductile packings. As the shear band broadens, more particles actively participate to the training. The existence of a shear band is thereby necessary to store a memory by cyclic shear in brittle packings, suggesting that memories can only be formed in marginally stable regions of the system.

\textit{Conclusions} --
In this Letter, we explore the role of mechanical stability in the context of memory training by cyclic shear in jammed solids. While marginally stable packings are able to store memories for all the explored strain amplitudes, we observe that highly stable packings need to first overcome brittle yielding and form a shear band in order to do so. Here is where mechanical stability comes into play: brittle packings become marginally stable after yielding and marginal stability is confined in the shear band where most of the rearrangements during the training take place. This result shows that memory training in jammed packings is only possible if the system, or a portion of it, is marginally stable.

The strong connection between memory training and mechanical stability suggests that the development of memory in real space is coupled to the evolution of the low-frequency vibrational modes, an aspect of memory training which requires further investigation. An exciting new direction would be to extend the work conducted here to soft sphere packings driven by athermal quasi-static random displacements, an active matter model introduced in theory~\cite{agoritsas_mean-field_2021} and simulations~\cite{morse_direct_2021}, where the brittle failure happens in regions randomly distributed across the system. Training a highly stable packing with this new cyclic driving could potentially allow for encoding memories in pockets of the system which could be preemptively designed, broadening the application scope of trainable jammed solids.

We thank Ian Graham, Peter Morse, Sidney Nagel and Sri Sastry for fruitful discussions on memory formation. This work was funded by the NSF Career Award grant No. DMR-1255370, and the Simons Collaboration on Cracking the Glass Problem (No. 454939 E. Corwin).

\bibliography{shearMemory.bib}

\begin{thebibliography}{30}%
\makeatletter
\providecommand \@ifxundefined [1]{%
 \@ifx{#1\undefined}
}%
\providecommand \@ifnum [1]{%
 \ifnum #1\expandafter \@firstoftwo
 \else \expandafter \@secondoftwo
 \fi
}%
\providecommand \@ifx [1]{%
 \ifx #1\expandafter \@firstoftwo
 \else \expandafter \@secondoftwo
 \fi
}%
\providecommand \natexlab [1]{#1}%
\providecommand \enquote  [1]{``#1''}%
\providecommand \bibnamefont  [1]{#1}%
\providecommand \bibfnamefont [1]{#1}%
\providecommand \citenamefont [1]{#1}%
\providecommand \href@noop [0]{\@secondoftwo}%
\providecommand \href [0]{\begingroup \@sanitize@url \@href}%
\providecommand \@href[1]{\@@startlink{#1}\@@href}%
\providecommand \@@href[1]{\endgroup#1\@@endlink}%
\providecommand \@sanitize@url [0]{\catcode `\\12\catcode `\$12\catcode
  `\&12\catcode `\#12\catcode `\^12\catcode `\_12\catcode `\%12\relax}%
\providecommand \@@startlink[1]{}%
\providecommand \@@endlink[0]{}%
\providecommand \url  [0]{\begingroup\@sanitize@url \@url }%
\providecommand \@url [1]{\endgroup\@href {#1}{\urlprefix }}%
\providecommand \urlprefix  [0]{URL }%
\providecommand \Eprint [0]{\href }%
\providecommand \doibase [0]{http://dx.doi.org/}%
\providecommand \selectlanguage [0]{\@gobble}%
\providecommand \bibinfo  [0]{\@secondoftwo}%
\providecommand \bibfield  [0]{\@secondoftwo}%
\providecommand \translation [1]{[#1]}%
\providecommand \BibitemOpen [0]{}%
\providecommand \bibitemStop [0]{}%
\providecommand \bibitemNoStop [0]{.\EOS\space}%
\providecommand \EOS [0]{\spacefactor3000\relax}%
\providecommand \BibitemShut  [1]{\csname bibitem#1\endcsname}%
\let\auto@bib@innerbib\@empty
\bibitem [{\citenamefont {Corté}\ \emph {et~al.}(2008)\citenamefont {Corté},
  \citenamefont {Chaikin}, \citenamefont {Gollub},\ and\ \citenamefont
  {Pine}}]{corte_random_2008}%
  \BibitemOpen
  \bibfield  {author} {\bibinfo {author} {\bibfnamefont {Laurent}\ \bibnamefont
  {Corté}}, \bibinfo {author} {\bibfnamefont {P.~M.}\ \bibnamefont {Chaikin}},
  \bibinfo {author} {\bibfnamefont {J.~P.}\ \bibnamefont {Gollub}}, \ and\
  \bibinfo {author} {\bibfnamefont {D.~J.}\ \bibnamefont {Pine}},\ }\bibfield
  {title} {{\selectlanguage {english}\enquote {\bibinfo {title} {Random
  organization in periodically driven systems},}\ }}\href {\doibase
  10.1038/nphys891} {\bibfield  {journal} {\bibinfo  {journal} {Nature
  Physics}\ }\textbf {\bibinfo {volume} {4}},\ \bibinfo {pages} {420--424}
  (\bibinfo {year} {2008})},\ \bibinfo {note} {number: 5 Publisher: Nature
  Publishing Group}\BibitemShut {NoStop}%
\bibitem [{\citenamefont {Fiocco}\ \emph {et~al.}(2014)\citenamefont {Fiocco},
  \citenamefont {Foffi},\ and\ \citenamefont {Sastry}}]{fiocco_encoding_2014}%
  \BibitemOpen
  \bibfield  {author} {\bibinfo {author} {\bibfnamefont {Davide}\ \bibnamefont
  {Fiocco}}, \bibinfo {author} {\bibfnamefont {Giuseppe}\ \bibnamefont
  {Foffi}}, \ and\ \bibinfo {author} {\bibfnamefont {Srikanth}\ \bibnamefont
  {Sastry}},\ }\bibfield  {title} {{\selectlanguage {english}\enquote {\bibinfo
  {title} {Encoding of {Memory} in {Sheared} {Amorphous} {Solids}},}\ }}\href
  {\doibase 10.1103/PhysRevLett.112.025702} {\bibfield  {journal} {\bibinfo
  {journal} {Physical Review Letters}\ }\textbf {\bibinfo {volume} {112}}
  (\bibinfo {year} {2014}),\ 10.1103/PhysRevLett.112.025702}\BibitemShut
  {NoStop}%
\bibitem [{\citenamefont {Keim}\ \emph {et~al.}(2019)\citenamefont {Keim},
  \citenamefont {Paulsen}, \citenamefont {Zeravcic}, \citenamefont {Sastry},\
  and\ \citenamefont {Nagel}}]{keim_memory_2019}%
  \BibitemOpen
  \bibfield  {author} {\bibinfo {author} {\bibfnamefont {Nathan~C.}\
  \bibnamefont {Keim}}, \bibinfo {author} {\bibfnamefont {Joseph~D.}\
  \bibnamefont {Paulsen}}, \bibinfo {author} {\bibfnamefont {Zorana}\
  \bibnamefont {Zeravcic}}, \bibinfo {author} {\bibfnamefont {Srikanth}\
  \bibnamefont {Sastry}}, \ and\ \bibinfo {author} {\bibfnamefont {Sidney~R.}\
  \bibnamefont {Nagel}},\ }\bibfield  {title} {{\selectlanguage
  {english}\enquote {\bibinfo {title} {Memory formation in matter},}\ }}\href
  {\doibase 10.1103/RevModPhys.91.035002} {\bibfield  {journal} {\bibinfo
  {journal} {Reviews of Modern Physics}\ }\textbf {\bibinfo {volume} {91}}
  (\bibinfo {year} {2019}),\ 10.1103/RevModPhys.91.035002}\BibitemShut
  {NoStop}%
\bibitem [{\citenamefont {Fiocco}\ \emph {et~al.}(2013)\citenamefont {Fiocco},
  \citenamefont {Foffi},\ and\ \citenamefont
  {Sastry}}]{fiocco_oscillatory_2013}%
  \BibitemOpen
  \bibfield  {author} {\bibinfo {author} {\bibfnamefont {Davide}\ \bibnamefont
  {Fiocco}}, \bibinfo {author} {\bibfnamefont {Giuseppe}\ \bibnamefont
  {Foffi}}, \ and\ \bibinfo {author} {\bibfnamefont {Srikanth}\ \bibnamefont
  {Sastry}},\ }\bibfield  {title} {{\selectlanguage {english}\enquote {\bibinfo
  {title} {Oscillatory athermal quasistatic deformation of a model glass},}\
  }}\href {\doibase 10.1103/PhysRevE.88.020301} {\bibfield  {journal} {\bibinfo
   {journal} {Physical Review E}\ }\textbf {\bibinfo {volume} {88}} (\bibinfo
  {year} {2013}),\ 10.1103/PhysRevE.88.020301}\BibitemShut {NoStop}%
\bibitem [{\citenamefont {Charbonneau}\ \emph {et~al.}(2014)\citenamefont
  {Charbonneau}, \citenamefont {Kurchan}, \citenamefont {Parisi}, \citenamefont
  {Urbani},\ and\ \citenamefont {Zamponi}}]{charbonneau_fractal_2014-4}%
  \BibitemOpen
  \bibfield  {author} {\bibinfo {author} {\bibfnamefont {Patrick}\ \bibnamefont
  {Charbonneau}}, \bibinfo {author} {\bibfnamefont {Jorge}\ \bibnamefont
  {Kurchan}}, \bibinfo {author} {\bibfnamefont {Giorgio}\ \bibnamefont
  {Parisi}}, \bibinfo {author} {\bibfnamefont {Pierfrancesco}\ \bibnamefont
  {Urbani}}, \ and\ \bibinfo {author} {\bibfnamefont {Francesco}\ \bibnamefont
  {Zamponi}},\ }\bibfield  {title} {{\selectlanguage {english}\enquote
  {\bibinfo {title} {Fractal free energy landscapes in structural glasses},}\
  }}\href {\doibase 10.1038/ncomms4725} {\bibfield  {journal} {\bibinfo
  {journal} {Nature Communications}\ }\textbf {\bibinfo {volume} {5}} (\bibinfo
  {year} {2014}),\ 10.1038/ncomms4725}\BibitemShut {NoStop}%
\bibitem [{\citenamefont {Jin}\ and\ \citenamefont
  {Yoshino}(2017)}]{jin_exploring_2017}%
  \BibitemOpen
  \bibfield  {author} {\bibinfo {author} {\bibfnamefont {Yuliang}\ \bibnamefont
  {Jin}}\ and\ \bibinfo {author} {\bibfnamefont {Hajime}\ \bibnamefont
  {Yoshino}},\ }\bibfield  {title} {{\selectlanguage {english}\enquote
  {\bibinfo {title} {Exploring the complex free-energy landscape of the
  simplest glass by rheology},}\ }}\href {\doibase 10.1038/ncomms14935}
  {\bibfield  {journal} {\bibinfo  {journal} {Nat Commun}\ }\textbf {\bibinfo
  {volume} {8}},\ \bibinfo {pages} {14935} (\bibinfo {year}
  {2017})}\BibitemShut {NoStop}%
\bibitem [{\citenamefont {Szulc}\ \emph {et~al.}(2020)\citenamefont {Szulc},
  \citenamefont {Gat},\ and\ \citenamefont {Regev}}]{szulc_forced_2020}%
  \BibitemOpen
  \bibfield  {author} {\bibinfo {author} {\bibfnamefont {Asaf}\ \bibnamefont
  {Szulc}}, \bibinfo {author} {\bibfnamefont {Omri}\ \bibnamefont {Gat}}, \
  and\ \bibinfo {author} {\bibfnamefont {Ido}\ \bibnamefont {Regev}},\
  }\bibfield  {title} {{\selectlanguage {english}\enquote {\bibinfo {title}
  {Forced deterministic dynamics on a random energy landscape: {Implications}
  for the physics of amorphous solids},}\ }}\href {\doibase
  10.1103/PhysRevE.101.052616} {\bibfield  {journal} {\bibinfo  {journal}
  {Phys. Rev. E}\ }\textbf {\bibinfo {volume} {101}},\ \bibinfo {pages}
  {052616} (\bibinfo {year} {2020})}\BibitemShut {NoStop}%
\bibitem [{\citenamefont {Adhikari}\ and\ \citenamefont
  {Sastry}(2018)}]{adhikari_memory_2018}%
  \BibitemOpen
  \bibfield  {author} {\bibinfo {author} {\bibfnamefont {Monoj}\ \bibnamefont
  {Adhikari}}\ and\ \bibinfo {author} {\bibfnamefont {Srikanth}\ \bibnamefont
  {Sastry}},\ }\bibfield  {title} {{\selectlanguage {english}\enquote {\bibinfo
  {title} {Memory formation in cyclically deformed amorphous solids and sphere
  assemblies},}\ }}\href {\doibase 10.1140/epje/i2018-11717-5} {\bibfield
  {journal} {\bibinfo  {journal} {Eur. Phys. J. E}\ }\textbf {\bibinfo {volume}
  {41}},\ \bibinfo {pages} {105} (\bibinfo {year} {2018})}\BibitemShut
  {NoStop}%
\bibitem [{\citenamefont {Ninarello}\ \emph {et~al.}(2017)\citenamefont
  {Ninarello}, \citenamefont {Berthier},\ and\ \citenamefont
  {Coslovich}}]{ninarello_models_2017}%
  \BibitemOpen
  \bibfield  {author} {\bibinfo {author} {\bibfnamefont {Andrea}\ \bibnamefont
  {Ninarello}}, \bibinfo {author} {\bibfnamefont {Ludovic}\ \bibnamefont
  {Berthier}}, \ and\ \bibinfo {author} {\bibfnamefont {Daniele}\ \bibnamefont
  {Coslovich}},\ }\bibfield  {title} {{\selectlanguage {english}\enquote
  {\bibinfo {title} {Models and {Algorithms} for the {Next} {Generation} of
  {Glass} {Transition} {Studies}},}\ }}\href {\doibase
  10.1103/PhysRevX.7.021039} {\bibfield  {journal} {\bibinfo  {journal} {Phys.
  Rev. X}\ }\textbf {\bibinfo {volume} {7}},\ \bibinfo {pages} {021039}
  (\bibinfo {year} {2017})}\BibitemShut {NoStop}%
\bibitem [{\citenamefont {Kapteijns}\ \emph {et~al.}(2019)\citenamefont
  {Kapteijns}, \citenamefont {Ji}, \citenamefont {Brito}, \citenamefont
  {Wyart},\ and\ \citenamefont {Lerner}}]{kapteijns_fast_2019}%
  \BibitemOpen
  \bibfield  {author} {\bibinfo {author} {\bibfnamefont {Geert}\ \bibnamefont
  {Kapteijns}}, \bibinfo {author} {\bibfnamefont {Wencheng}\ \bibnamefont
  {Ji}}, \bibinfo {author} {\bibfnamefont {Carolina}\ \bibnamefont {Brito}},
  \bibinfo {author} {\bibfnamefont {Matthieu}\ \bibnamefont {Wyart}}, \ and\
  \bibinfo {author} {\bibfnamefont {Edan}\ \bibnamefont {Lerner}},\ }\bibfield
  {title} {{\selectlanguage {english}\enquote {\bibinfo {title} {Fast
  generation of ultrastable computer glasses by minimization of an augmented
  potential energy},}\ }}\href {\doibase 10.1103/PhysRevE.99.012106} {\bibfield
   {journal} {\bibinfo  {journal} {Phys. Rev. E}\ }\textbf {\bibinfo {volume}
  {99}},\ \bibinfo {pages} {012106} (\bibinfo {year} {2019})},\ \bibinfo {note}
  {publisher: American Physical Society}\BibitemShut {NoStop}%
\bibitem [{\citenamefont {Rainone}\ and\ \citenamefont
  {Urbani}(2016)}]{rainone_following_2016-1}%
  \BibitemOpen
  \bibfield  {author} {\bibinfo {author} {\bibfnamefont {Corrado}\ \bibnamefont
  {Rainone}}\ and\ \bibinfo {author} {\bibfnamefont {Pierfrancesco}\
  \bibnamefont {Urbani}},\ }\bibfield  {title} {{\selectlanguage
  {english}\enquote {\bibinfo {title} {Following the evolution of glassy states
  under external perturbations: the full replica symmetry breaking solution},}\
  }}\href {\doibase 10.1088/1742-5468/2016/05/053302} {\bibfield  {journal}
  {\bibinfo  {journal} {J. Stat. Mech.}\ }\textbf {\bibinfo {volume} {2016}},\
  \bibinfo {pages} {053302} (\bibinfo {year} {2016})}\BibitemShut {NoStop}%
\bibitem [{\citenamefont {Jin}\ \emph {et~al.}(2018)\citenamefont {Jin},
  \citenamefont {Urbani}, \citenamefont {Zamponi},\ and\ \citenamefont
  {Yoshino}}]{jin_stability-reversibility_2018}%
  \BibitemOpen
  \bibfield  {author} {\bibinfo {author} {\bibfnamefont {Yuliang}\ \bibnamefont
  {Jin}}, \bibinfo {author} {\bibfnamefont {Pierfrancesco}\ \bibnamefont
  {Urbani}}, \bibinfo {author} {\bibfnamefont {Francesco}\ \bibnamefont
  {Zamponi}}, \ and\ \bibinfo {author} {\bibfnamefont {Hajime}\ \bibnamefont
  {Yoshino}},\ }\bibfield  {title} {{\selectlanguage {english}\enquote
  {\bibinfo {title} {A stability-reversibility map unifies elasticity,
  plasticity, yielding, and jamming in hard sphere glasses},}\ }}\href
  {\doibase 10.1126/sciadv.aat6387} {\bibfield  {journal} {\bibinfo  {journal}
  {Sci. Adv.}\ }\textbf {\bibinfo {volume} {4}},\ \bibinfo {pages} {eaat6387}
  (\bibinfo {year} {2018})}\BibitemShut {NoStop}%
\bibitem [{\citenamefont {Yeh}\ \emph {et~al.}(2020)\citenamefont {Yeh},
  \citenamefont {Ozawa}, \citenamefont {Miyazaki}, \citenamefont {Kawasaki},\
  and\ \citenamefont {Berthier}}]{yeh_glass_2020}%
  \BibitemOpen
  \bibfield  {author} {\bibinfo {author} {\bibfnamefont {Wei-Ting}\
  \bibnamefont {Yeh}}, \bibinfo {author} {\bibfnamefont {Misaki}\ \bibnamefont
  {Ozawa}}, \bibinfo {author} {\bibfnamefont {Kunimasa}\ \bibnamefont
  {Miyazaki}}, \bibinfo {author} {\bibfnamefont {Takeshi}\ \bibnamefont
  {Kawasaki}}, \ and\ \bibinfo {author} {\bibfnamefont {Ludovic}\ \bibnamefont
  {Berthier}},\ }\bibfield  {title} {{\selectlanguage {english}\enquote
  {\bibinfo {title} {Glass {Stability} {Changes} the {Nature} of {Yielding}
  under {Oscillatory} {Shear}},}\ }}\href {\doibase
  10.1103/PhysRevLett.124.225502} {\bibfield  {journal} {\bibinfo  {journal}
  {Physical Review Letters}\ }\textbf {\bibinfo {volume} {124}} (\bibinfo
  {year} {2020}),\ 10.1103/PhysRevLett.124.225502}\BibitemShut {NoStop}%
\bibitem [{\citenamefont {Biroli}\ and\ \citenamefont
  {Urbani}(2016)}]{biroli_breakdown_2016-1}%
  \BibitemOpen
  \bibfield  {author} {\bibinfo {author} {\bibfnamefont {Giulio}\ \bibnamefont
  {Biroli}}\ and\ \bibinfo {author} {\bibfnamefont {Pierfrancesco}\
  \bibnamefont {Urbani}},\ }\bibfield  {title} {{\selectlanguage
  {english}\enquote {\bibinfo {title} {Breakdown of elasticity in amorphous
  solids},}\ }}\href {\doibase 10.1038/nphys3845} {\bibfield  {journal}
  {\bibinfo  {journal} {Nature Phys}\ }\textbf {\bibinfo {volume} {12}},\
  \bibinfo {pages} {1130--1133} (\bibinfo {year} {2016})}\BibitemShut {NoStop}%
\bibitem [{\citenamefont {Charbonneau}\ \emph {et~al.}(2017)\citenamefont
  {Charbonneau}, \citenamefont {Kurchan}, \citenamefont {Parisi}, \citenamefont
  {Urbani},\ and\ \citenamefont {Zamponi}}]{charbonneau_glass_2017-3}%
  \BibitemOpen
  \bibfield  {author} {\bibinfo {author} {\bibfnamefont {Patrick}\ \bibnamefont
  {Charbonneau}}, \bibinfo {author} {\bibfnamefont {Jorge}\ \bibnamefont
  {Kurchan}}, \bibinfo {author} {\bibfnamefont {Giorgio}\ \bibnamefont
  {Parisi}}, \bibinfo {author} {\bibfnamefont {Pierfrancesco}\ \bibnamefont
  {Urbani}}, \ and\ \bibinfo {author} {\bibfnamefont {Francesco}\ \bibnamefont
  {Zamponi}},\ }\bibfield  {title} {{\selectlanguage {english}\enquote
  {\bibinfo {title} {Glass and {Jamming} {Transitions}: {From} {Exact}
  {Results} to {Finite}-{Dimensional} {Descriptions}},}\ }}\href {\doibase
  10.1146/annurev-conmatphys-031016-025334} {\bibfield  {journal} {\bibinfo
  {journal} {Annual Review of Condensed Matter Physics}\ }\textbf {\bibinfo
  {volume} {8}},\ \bibinfo {pages} {265--288} (\bibinfo {year}
  {2017})}\BibitemShut {NoStop}%
\bibitem [{\citenamefont {Hagh}\ \emph {et~al.}(2021)\citenamefont {Hagh},
  \citenamefont {Nage}, \citenamefont {Liu}, \citenamefont {Manning},\ and\
  \citenamefont {Corwin}}]{hagh_transient_2021}%
  \BibitemOpen
  \bibfield  {author} {\bibinfo {author} {\bibfnamefont {Varda~F.}\
  \bibnamefont {Hagh}}, \bibinfo {author} {\bibfnamefont {Sidney~R.}\
  \bibnamefont {Nage}}, \bibinfo {author} {\bibfnamefont {Andrea~J.}\
  \bibnamefont {Liu}}, \bibinfo {author} {\bibfnamefont {M.~Lisa}\ \bibnamefont
  {Manning}}, \ and\ \bibinfo {author} {\bibfnamefont {Eric~I.}\ \bibnamefont
  {Corwin}},\ }\bibfield  {title} {{\selectlanguage {english}\enquote {\bibinfo
  {title} {Transient degrees of freedom and stability},}\ }}\href
  {http://arxiv.org/abs/2105.10846} {\bibfield  {journal} {\bibinfo  {journal}
  {arXiv:2105.10846 [cond-mat]}\ } (\bibinfo {year} {2021})},\ \bibinfo {note}
  {arXiv: 2105.10846}\BibitemShut {NoStop}%
\bibitem [{\citenamefont {Durian}(1995)}]{durian_foam_1995-1}%
  \BibitemOpen
  \bibfield  {author} {\bibinfo {author} {\bibfnamefont {D.~J.}\ \bibnamefont
  {Durian}},\ }\bibfield  {title} {{\selectlanguage {english}\enquote {\bibinfo
  {title} {Foam {Mechanics} at the {Bubble} {Scale}},}\ }}\href {\doibase
  10.1103/PhysRevLett.75.4780} {\bibfield  {journal} {\bibinfo  {journal}
  {Physical Review Letters}\ }\textbf {\bibinfo {volume} {75}},\ \bibinfo
  {pages} {4780--4783} (\bibinfo {year} {1995})}\BibitemShut {NoStop}%
\bibitem [{\citenamefont {O’Hern}\ \emph {et~al.}(2003)\citenamefont
  {O’Hern}, \citenamefont {Silbert}, \citenamefont {Liu},\ and\ \citenamefont
  {Nagel}}]{ohern_jamming_2003-3}%
  \BibitemOpen
  \bibfield  {author} {\bibinfo {author} {\bibfnamefont {Corey~S.}\
  \bibnamefont {O’Hern}}, \bibinfo {author} {\bibfnamefont {Leonardo~E.}\
  \bibnamefont {Silbert}}, \bibinfo {author} {\bibfnamefont {Andrea~J.}\
  \bibnamefont {Liu}}, \ and\ \bibinfo {author} {\bibfnamefont {Sidney~R.}\
  \bibnamefont {Nagel}},\ }\bibfield  {title} {{\selectlanguage
  {english}\enquote {\bibinfo {title} {Jamming at zero temperature and zero
  applied stress: {The} epitome of disorder},}\ }}\href {\doibase
  10.1103/PhysRevE.68.011306} {\bibfield  {journal} {\bibinfo  {journal} {Phys.
  Rev. E}\ }\textbf {\bibinfo {volume} {68}},\ \bibinfo {pages} {011306}
  (\bibinfo {year} {2003})}\BibitemShut {NoStop}%
\bibitem [{\citenamefont {Maloney}\ and\ \citenamefont
  {Lemaître}(2006)}]{maloney_amorphous_2006-2}%
  \BibitemOpen
  \bibfield  {author} {\bibinfo {author} {\bibfnamefont {Craig~E.}\
  \bibnamefont {Maloney}}\ and\ \bibinfo {author} {\bibfnamefont {Anaël}\
  \bibnamefont {Lemaître}},\ }\bibfield  {title} {{\selectlanguage
  {english}\enquote {\bibinfo {title} {Amorphous systems in athermal,
  quasistatic shear},}\ }}\href {\doibase 10.1103/PhysRevE.74.016118}
  {\bibfield  {journal} {\bibinfo  {journal} {Phys. Rev. E}\ }\textbf {\bibinfo
  {volume} {74}},\ \bibinfo {pages} {016118} (\bibinfo {year} {2006})},\
  \bibinfo {note} {publisher: American Physical Society}\BibitemShut {NoStop}%
\bibitem [{\citenamefont {Maloney}\ and\ \citenamefont
  {Lemaître}(2004)}]{maloney_universal_2004}%
  \BibitemOpen
  \bibfield  {author} {\bibinfo {author} {\bibfnamefont {Craig}\ \bibnamefont
  {Maloney}}\ and\ \bibinfo {author} {\bibfnamefont {Anaël}\ \bibnamefont
  {Lemaître}},\ }\bibfield  {title} {{\selectlanguage {english}\enquote
  {\bibinfo {title} {Universal {Breakdown} of {Elasticity} at the {Onset} of
  {Material} {Failure}},}\ }}\href {\doibase 10.1103/PhysRevLett.93.195501}
  {\bibfield  {journal} {\bibinfo  {journal} {Phys. Rev. Lett.}\ }\textbf
  {\bibinfo {volume} {93}},\ \bibinfo {pages} {195501} (\bibinfo {year}
  {2004})},\ \bibinfo {note} {publisher: American Physical Society}\BibitemShut
  {NoStop}%
\bibitem [{\citenamefont {Ozawa}\ \emph {et~al.}(2018)\citenamefont {Ozawa},
  \citenamefont {Berthier}, \citenamefont {Biroli}, \citenamefont {Rosso},\
  and\ \citenamefont {Tarjus}}]{ozawa_random_2018}%
  \BibitemOpen
  \bibfield  {author} {\bibinfo {author} {\bibfnamefont {Misaki}\ \bibnamefont
  {Ozawa}}, \bibinfo {author} {\bibfnamefont {Ludovic}\ \bibnamefont
  {Berthier}}, \bibinfo {author} {\bibfnamefont {Giulio}\ \bibnamefont
  {Biroli}}, \bibinfo {author} {\bibfnamefont {Alberto}\ \bibnamefont {Rosso}},
  \ and\ \bibinfo {author} {\bibfnamefont {Gilles}\ \bibnamefont {Tarjus}},\
  }\bibfield  {title} {{\selectlanguage {english}\enquote {\bibinfo {title}
  {Random critical point separates brittle and ductile yielding transitions in
  amorphous materials},}\ }}\href {\doibase 10.1073/pnas.1806156115} {\bibfield
   {journal} {\bibinfo  {journal} {PNAS}\ }\textbf {\bibinfo {volume} {115}},\
  \bibinfo {pages} {6656--6661} (\bibinfo {year} {2018})}\BibitemShut {NoStop}%
\bibitem [{\citenamefont {Morse}\ and\ \citenamefont
  {Corwin}(2014)}]{morse_geometric_2014-5}%
  \BibitemOpen
  \bibfield  {author} {\bibinfo {author} {\bibfnamefont {Peter~K.}\
  \bibnamefont {Morse}}\ and\ \bibinfo {author} {\bibfnamefont {Eric~I.}\
  \bibnamefont {Corwin}},\ }\bibfield  {title} {{\selectlanguage
  {english}\enquote {\bibinfo {title} {Geometric {Signatures} of {Jamming} in
  the {Mechanical} {Vacuum}},}\ }}\href {\doibase
  10.1103/PhysRevLett.112.115701} {\bibfield  {journal} {\bibinfo  {journal}
  {Physical Review Letters}\ }\textbf {\bibinfo {volume} {112}} (\bibinfo
  {year} {2014}),\ 10.1103/PhysRevLett.112.115701}\BibitemShut {NoStop}%
\bibitem [{\citenamefont {Charbonneau}\ \emph {et~al.}(2015)\citenamefont
  {Charbonneau}, \citenamefont {Corwin}, \citenamefont {Parisi},\ and\
  \citenamefont {Zamponi}}]{charbonneau_jamming_2015-6}%
  \BibitemOpen
  \bibfield  {author} {\bibinfo {author} {\bibfnamefont {Patrick}\ \bibnamefont
  {Charbonneau}}, \bibinfo {author} {\bibfnamefont {Eric~I.}\ \bibnamefont
  {Corwin}}, \bibinfo {author} {\bibfnamefont {Giorgio}\ \bibnamefont
  {Parisi}}, \ and\ \bibinfo {author} {\bibfnamefont {Francesco}\ \bibnamefont
  {Zamponi}},\ }\bibfield  {title} {{\selectlanguage {english}\enquote
  {\bibinfo {title} {Jamming {Criticality} {Revealed} by {Removing} {Localized}
  {Buckling} {Excitations}},}\ }}\href {\doibase
  10.1103/PhysRevLett.114.125504} {\bibfield  {journal} {\bibinfo  {journal}
  {Physical Review Letters}\ }\textbf {\bibinfo {volume} {114}} (\bibinfo
  {year} {2015}),\ 10.1103/PhysRevLett.114.125504}\BibitemShut {NoStop}%
\bibitem [{\citenamefont {Bitzek}\ \emph {et~al.}(2006)\citenamefont {Bitzek},
  \citenamefont {Koskinen}, \citenamefont {Gähler}, \citenamefont {Moseler},\
  and\ \citenamefont {Gumbsch}}]{bitzek_structural_2006-2}%
  \BibitemOpen
  \bibfield  {author} {\bibinfo {author} {\bibfnamefont {Erik}\ \bibnamefont
  {Bitzek}}, \bibinfo {author} {\bibfnamefont {Pekka}\ \bibnamefont
  {Koskinen}}, \bibinfo {author} {\bibfnamefont {Franz}\ \bibnamefont
  {Gähler}}, \bibinfo {author} {\bibfnamefont {Michael}\ \bibnamefont
  {Moseler}}, \ and\ \bibinfo {author} {\bibfnamefont {Peter}\ \bibnamefont
  {Gumbsch}},\ }\bibfield  {title} {{\selectlanguage {english}\enquote
  {\bibinfo {title} {Structural {Relaxation} {Made} {Simple}},}\ }}\href
  {\doibase 10.1103/PhysRevLett.97.170201} {\bibfield  {journal} {\bibinfo
  {journal} {Phys. Rev. Lett.}\ }\textbf {\bibinfo {volume} {97}},\ \bibinfo
  {pages} {170201} (\bibinfo {year} {2006})}\BibitemShut {NoStop}%
\bibitem [{\citenamefont {Manning}\ and\ \citenamefont
  {Liu}(2011)}]{manning_vibrational_2011-2}%
  \BibitemOpen
  \bibfield  {author} {\bibinfo {author} {\bibfnamefont {M.~L.}\ \bibnamefont
  {Manning}}\ and\ \bibinfo {author} {\bibfnamefont {A.~J.}\ \bibnamefont
  {Liu}},\ }\bibfield  {title} {{\selectlanguage {english}\enquote {\bibinfo
  {title} {Vibrational {Modes} {Identify} {Soft} {Spots} in a {Sheared}
  {Disordered} {Packing}},}\ }}\href {\doibase 10.1103/PhysRevLett.107.108302}
  {\bibfield  {journal} {\bibinfo  {journal} {Phys. Rev. Lett.}\ }\textbf
  {\bibinfo {volume} {107}},\ \bibinfo {pages} {108302} (\bibinfo {year}
  {2011})},\ \bibinfo {note} {publisher: American Physical Society}\BibitemShut
  {NoStop}%
\bibitem [{\citenamefont {Mizuno}\ \emph {et~al.}(2017)\citenamefont {Mizuno},
  \citenamefont {Shiba},\ and\ \citenamefont {Ikeda}}]{mizuno_continuum_2017}%
  \BibitemOpen
  \bibfield  {author} {\bibinfo {author} {\bibfnamefont {Hideyuki}\
  \bibnamefont {Mizuno}}, \bibinfo {author} {\bibfnamefont {Hayato}\
  \bibnamefont {Shiba}}, \ and\ \bibinfo {author} {\bibfnamefont {Atsushi}\
  \bibnamefont {Ikeda}},\ }\bibfield  {title} {{\selectlanguage
  {English}\enquote {\bibinfo {title} {Continuum limit of the vibrational
  properties of amorphous solids},}\ }}\href {\doibase 10.1073/pnas.1709015114}
  {\bibfield  {journal} {\bibinfo  {journal} {Proceedings of the National
  Academy of Sciences}\ }\textbf {\bibinfo {volume} {114}},\ \bibinfo {pages}
  {E9767--E9774} (\bibinfo {year} {2017})}\BibitemShut {NoStop}%
\bibitem [{\citenamefont {Keim}\ and\ \citenamefont
  {Nagel}(2011)}]{keim_generic_2011}%
  \BibitemOpen
  \bibfield  {author} {\bibinfo {author} {\bibfnamefont {Nathan~C.}\
  \bibnamefont {Keim}}\ and\ \bibinfo {author} {\bibfnamefont {Sidney~R.}\
  \bibnamefont {Nagel}},\ }\bibfield  {title} {{\selectlanguage
  {english}\enquote {\bibinfo {title} {Generic {Transient} {Memory} {Formation}
  in {Disordered} {Systems} with {Noise}},}\ }}\href {\doibase
  10.1103/PhysRevLett.107.010603} {\bibfield  {journal} {\bibinfo  {journal}
  {Physical Review Letters}\ }\textbf {\bibinfo {volume} {107}} (\bibinfo
  {year} {2011}),\ 10.1103/PhysRevLett.107.010603}\BibitemShut {NoStop}%
\bibitem [{\citenamefont {Goodrich}\ \emph {et~al.}(2012)\citenamefont
  {Goodrich}, \citenamefont {Liu},\ and\ \citenamefont
  {Nagel}}]{goodrich_finite-size_2012-5}%
  \BibitemOpen
  \bibfield  {author} {\bibinfo {author} {\bibfnamefont {Carl~P.}\ \bibnamefont
  {Goodrich}}, \bibinfo {author} {\bibfnamefont {Andrea~J.}\ \bibnamefont
  {Liu}}, \ and\ \bibinfo {author} {\bibfnamefont {Sidney~R.}\ \bibnamefont
  {Nagel}},\ }\bibfield  {title} {{\selectlanguage {english}\enquote {\bibinfo
  {title} {Finite-{Size} {Scaling} at the {Jamming} {Transition}},}\ }}\href
  {\doibase 10.1103/PhysRevLett.109.095704} {\bibfield  {journal} {\bibinfo
  {journal} {Physical Review Letters}\ }\textbf {\bibinfo {volume} {109}}
  (\bibinfo {year} {2012}),\ 10.1103/PhysRevLett.109.095704}\BibitemShut
  {NoStop}%
\bibitem [{\citenamefont {Agoritsas}(2021)}]{agoritsas_mean-field_2021}%
  \BibitemOpen
  \bibfield  {author} {\bibinfo {author} {\bibfnamefont {Elisabeth}\
  \bibnamefont {Agoritsas}},\ }\bibfield  {title} {{\selectlanguage
  {english}\enquote {\bibinfo {title} {Mean-field dynamics of
  infinite-dimensional particle systems: global shear versus random local
  forcing},}\ }}\href {\doibase 10.1088/1742-5468/abdd18} {\bibfield  {journal}
  {\bibinfo  {journal} {J. Stat. Mech.}\ }\textbf {\bibinfo {volume} {2021}},\
  \bibinfo {pages} {033501} (\bibinfo {year} {2021})}\BibitemShut {NoStop}%
\bibitem [{\citenamefont {Morse}\ \emph {et~al.}(2021)\citenamefont {Morse},
  \citenamefont {Roy}, \citenamefont {Agoritsas}, \citenamefont {Stanifer},
  \citenamefont {Corwin},\ and\ \citenamefont {Manning}}]{morse_direct_2021}%
  \BibitemOpen
  \bibfield  {author} {\bibinfo {author} {\bibfnamefont {Peter~K.}\
  \bibnamefont {Morse}}, \bibinfo {author} {\bibfnamefont {Sudeshna}\
  \bibnamefont {Roy}}, \bibinfo {author} {\bibfnamefont {Elisabeth}\
  \bibnamefont {Agoritsas}}, \bibinfo {author} {\bibfnamefont {Ethan}\
  \bibnamefont {Stanifer}}, \bibinfo {author} {\bibfnamefont {Eric~I.}\
  \bibnamefont {Corwin}}, \ and\ \bibinfo {author} {\bibfnamefont {M.~Lisa}\
  \bibnamefont {Manning}},\ }\bibfield  {title} {{\selectlanguage
  {english}\enquote {\bibinfo {title} {A direct link between active matter and
  sheared granular systems},}\ }}\href {\doibase 10.1073/pnas.2019909118}
  {\bibfield  {journal} {\bibinfo  {journal} {Proc Natl Acad Sci USA}\ }\textbf
  {\bibinfo {volume} {118}},\ \bibinfo {pages} {e2019909118} (\bibinfo {year}
  {2021})}\BibitemShut {NoStop}%
\end{thebibliography}%

\end{document}